\documentclass[aps, singlecolumn, showpacs]{revtex4}

\begin{document}
\title{Role of local duality invariance in axion electrodynamics of topological insulators}
\author{S. C. Tiwari \\
Institute of Natural Philosophy \\
1 Kusum Kutir, Mahamanapuri \\
Varanasi 221005, India }
\begin{abstract}
Advances in material technology and confluence of ideas from particle physics, quantum field theory and condensed matter physics have led to the discovery of new states of matter as well as new physical phenomena: one of them termed as topological insulator has attracted a great deal of attention recently. Speculations on the possibility of observing the most elusive objects like axions and monopoles in topological insulators have led to studies that emphasize the role of symmetry and universality. In this paper we argue that electric-magnetic duality could be of deep significance in this context. We develop a duality invariant theory of topological insulators and show that under appropriate conditions this theory reduces to the axion electrodynamics for static case, topological quantization is related with the multi-valuedness of the duality gauge potential, and modifies Faraday's law for dynamical axion that would change the dispersion relation of axionic polariton. A new effect dual to the dynamical axion field effect is predicted and its physical consequences are discussed.
 
\end{abstract}
\maketitle

Spin-orbit coupling arises naturally in Dirac's relativistic equation of electron in external electromagnetic (EM) field \cite{1}; recent advances in the study of new materials have shown its renewed significance in condensed matter. One of the striking examples is the spin Hall effect predicted theoretically and soon established experimentally \cite{2, 3}. Spin Hall effect at a basic level is accumulation of the opposite spin states of electrons at the opposite edges of a material under suitable conditions of the application of EM field. Similar to quantum Hall effect it also gets quantized though the origin of quantization is different. Confluence of topology, symmetry (time reversal) and microphysical interaction mechanism (spin-orbit coupling) paved the path for the realization of so called 'exotic state of matter', namely topological insulator \cite{2, 3}. A simple way to understand topological insulator is to consider its response to the applied EM fields, first recalling the standard Maxwell theory in which for a linear isotropic medium the response at a macroscopic level is embodied in the constitutive relations
\begin{equation}
 {\bf D} = {\bf E} + 4 \pi {\bf P} = \epsilon {\bf E}
\end{equation}
\begin{equation}
 {\bf H} = {\bf B} - 4 \pi {\bf M} =\frac{ {\bf B}}{\mu}
\end{equation}
Here ${\bf P}$ and ${\bf M}$ are electric and magnetic polarization vectors, and the scalar quantities $\epsilon$ and $\mu$ are the usual permittivity and permeability respectively. A topological insulator (TI) is characterized by an additional parameter $\theta$ that couples to the pseudo-scalar product of electric and magnetic fields in the effective topological action \cite{2}
\begin{equation}
 S_{top} =\int \frac{\theta}{2\pi} \alpha {\bf E.B}~ d^4x
\end{equation}
and modifies the constitutive relations
\begin{equation}
 {\bf D} =\epsilon {\bf E} - \frac{\theta}{2\pi} \alpha {\bf B}
\end{equation}
\begin{equation}
 {\bf H} =\frac{\bf B}{\mu} + \frac{\theta}{2\pi} \alpha {\bf E}
\end{equation}
Here $\alpha = e^2/\hbar c$. A constant parameter $\theta$ implies that the integrand in (3) becomes a total derivative, therefore $S_{top}$ would not alter the field equations.

Note that the term ${\bf E.B}$ in (3) transforms as a pseudo-scalar under space reflection (P) and it is odd under time reversal (T); since the standard action has to be a Lorentz scalar $\theta$ must be odd under both P and T. Evidently in common insulators $\theta = 0$. Interesting possibility assuming T-invariance exists when $\theta = \pi$, and this defines the new class of insulators - TI. Authors in \cite{2} emphasize that the parameter $\theta$ corresponds to magneto-electric polarization in their theory and expressions (4) and (5) embody the topological magneto-electric effect (TME) in which magnetic(electric) field could induce electric(magnetic) polarization. Another new result is that an electric charge near the topological interface induces a mirror or image magnetic monopole \cite{4}. Intriguing generalization \cite{5}
is to break T symmetry via magnetic fluctuations so that $\theta$ becomes a dynamical field variable - these are called topological magnetic insulators (TMI).

Promoting $\theta$ to a dynamical field entails that the integrand in action (3) would change the field equation since it cannot be written in the form of a total derivative
\begin{equation}
 4 ~ {\bf E.B} = F_{\mu\nu} ~ ^*F^{\mu\nu} = \frac{1}{2}\partial^\mu (\epsilon_{\mu\nu\lambda\sigma} A^\nu \partial^\lambda A^\sigma)
\end{equation}
where EM field tensor $F_{\mu\nu} = \partial_\mu A_\nu -\partial_\nu A_\mu$ and its dual is obtained using totally anti-symmetric fourth rank Levi-Civita tensor $^*F_{\mu\nu} =\frac{1}{2} \epsilon_{\mu\nu\alpha\beta} F^{\alpha\beta}$. Euler-Lagrange equations in this case have formal resemblance with axion electrodynamics of particle physics: Wilczek in 1987 \cite{6} speculated that some of the axion field properties could possibly be realized in condensed matter system. This idea is nicely articulated for topological insulators in \cite{2} arguing that observing elusive objects axions and monopoles in laboratory seems feasible. Thus TI and TMI have wider ramifications extending to field theory and particle physics. 

Attention is drawn to another significant idea in the unification quest for fundamental interactions that of duality \cite{7}; it seems first work on duality in TI has been reported by Karch \cite{8} in which properties of TI are discussed in the framework of SL(2, Z) symmetry. The present paper is aimed at understanding generalized duality in EM field equations such that the rotation angle is a local space and time dependent variable. In this formulation it is found that axion electrodynamics appears as a special case under suitable conditions and topological quantization is related with the multi-valuedness of what we term as duality gauge potential. This theory predicts a new effect that could be interpreted as dual to the magnetic field induced electric charge in the axion electrodynamics \cite{6}. Local duality invariance gives new insights on the physics of TI as discussed in the following.

That source-free Maxwell equations in vacuum have a duality symmetry ${\bf E} \rightarrow {\bf B}$ and ${\bf B} \rightarrow -{\bf E}$ has been known since long \cite{7}. Postulating magnetic charge and current Maxwell equations in vacuum can be written in the covariant symmetric form
\begin{equation}
 \partial_\nu F^{\nu\mu} = 4 \pi J_e ^\mu
\end{equation}
\begin{equation}
 \partial_\nu ~ ^*F^{\nu\mu} = 4 \pi J_m ^\mu
\end{equation}
These equations are invariant under duality rotation with a real constant angle $\zeta$
\begin{equation}
 F_{\mu\nu} = F_{\mu\nu} cos\zeta + ^*F_{\mu\nu} sin\zeta
\end{equation}
\begin{equation}
 ^*F_{\mu\nu} = - F_{\mu\nu} sin \zeta + ^*F_{\mu\nu} cos \zeta
\end{equation} 
provided the source current densities also undergo similar rotation. One of the important applications of duality symmetry is that one could explain the absence of magnetic charges \cite{9}. Making a duality rotation with an angle such that the transformed $J_m^\mu = 0$ one gets the standard Maxwell equations: thus it is by convention that electron is assigned an electric charge $-e$ and a vanishing magnetic charge.

In a material medium duality rotation has to be applied carefully taking into account the constitutive relations (1) and (2) since the fields ${\bf D}$ and ${\bf H}$ would appear in the field equations \cite{8}. Here we give a different approach introducing antisymmetric polarization tensor $P^{\mu\nu}$ and construct its dual $^*P^{\mu\nu}$ using the Levi-Civita tensor. The components $P^{0i}$ and $P^{ij}$ represent the electric ${\bf P}$ and magnetic ${\bf M}$ $(P^{ij}=-\epsilon^{ijk} M_k)$ polarization vectors respectively. We can define the polarization current densities
\begin{equation}
J^\mu_P = \partial_\nu P^{\nu\mu}
\end{equation}
\begin{equation}
 ^*J^\mu_P = \partial_\nu ~ ^*P^{\nu\mu}
\end{equation}
and modify the Maxwell equations in material medium in the presence of sources in the following form
\begin{equation}
 \partial_\nu F^{\nu\mu} = 4 \pi J_e ^\mu + 4\pi J^\mu_P = 4\pi J^\mu
\end{equation}
\begin{equation}
\partial_\nu ~ ^*F^{\nu\mu} = 4 \pi J_m ^\mu + 4\pi ~ ^*J^\mu_P = 4\pi~ ^*J^\mu 
\end{equation}
Obviously Eqs (13) and (14) are invariant under the duality rotation assuming that the polarization current densities also transform in the same way as the source current densities. Invoking the argument that one could choose the rotation angle such that the transformed $^*J^\mu$ vanishes we obtain the Maxwell equations containing the macroscopic fields ${\bf D}$ and ${\bf H}$ in the traditional form.

It is now straightforward to implement local duality invariance in Maxwell equations making $\zeta$ to be an arbitrary function of space and time: a duality gauge 4-vector  $W_\nu$ has to be introduced that transforms as $W_\nu \rightarrow W_\nu + g^{-1} \partial_\nu \zeta$, and couples to the EM field tensor in the field equations with the coupling constant $g$
\begin{equation}
\partial_\nu F^{\nu\mu} = g W_\nu ~^*F^{\nu\mu} + 4\pi J^\mu
\end{equation}
\begin{equation}
\partial_\nu ~^*F^{\nu\mu} = - g W_\nu F^{\nu\mu} + 4\pi ~ ^*J^\mu
\end{equation}
Generalized Maxwell field equations (15) and (16) constitute one of the important results of this paper. This set of equations in vacuum and source-free case was obtained earlier using variational principle for a vector Lagrangian \cite{10}, however there are subtle technical aspects in connection with such a formulation for local duality discussed in detail elsewhere (in preparation). Here we work with the field equations and show that novel properties of topological insulators \cite{2, 3} could be understood based on them, and put forward the proposition that phenomenological physical mechanism for new electric and magnetic properties of matter, e. g. TI and TMI is related with the duality invariance.

Let us analyse the structure of Eqs (15) and (16): right hand sides comprise of three terms, namely duality induced term, polarization currents and true (free) charge currents. To understand them first we go back to Eqs. (13) and (14). In a usual way one can define new fields $G^{\mu\nu} = F^{\mu\nu} - P^{\mu\nu}$ and its dual and rewrite these equations in the form
\begin{equation}
\partial_\nu G^{\nu\mu} = 4\pi J^\mu_e 
\end{equation}
\begin{equation}
 \partial_\nu ~ ^*G^{\nu\mu} = 4\pi J^\mu_m 
\end{equation}
Alternatively since the polarization currents in a material originate due to bound charges i. e. electric dipole, magnetic dipole and multipoles, these could be clubbed together with the source currents as depicted in Eqs. (13) and (14). Thus we have two equivalent descriptions: modifying EM fields through constitutive relations as a material response as in Eqs. (17) and (18) and changing source currents in the presence of matter as in Eqs. (13) and (14). Extending this argument to Eqs. (15) and (16) the duality induced effect could be interpreted either in terms of altered constitutive relations or as a source of new current.

To make the physical interpretation more transparent let us first specialize to the case when $^*J^\mu_P = 0$, and assume simple relations (1) and (2) to write vector form of Eqs. (15) and (16)
\begin{equation}
 {\bf \nabla.D} = g {\bf W.B} +4\pi \rho_e
\end{equation}
\begin{equation}
 {\bf \nabla.B} = -g {\bf W.E} +4\pi \rho_m
\end{equation}
\begin{equation}
{\bf \nabla} \times {\bf E} + \frac {\partial {\bf B}}{\partial t} - g {\bf W} \times {\bf B} +g W_0 {\bf E} =4\pi {\bf J}_m
\end{equation}
\begin{equation}
 {\bf \nabla} \times {\bf H} - \frac {\partial {\bf D}}{\partial t} + g {\bf W} \times {\bf E} +g W_0 {\bf B} =4 \pi {\bf J}_e
\end{equation}

Conventionl assignment of discrete symmetry for electric charge density \cite{9} implies that ${\bf W}$ and $W_0$ are odd under both P and T transformations. We interpret $W_\mu$ as duality gauge field since it affects duality rotation, and seems to mediate spin interaction as reflected in the form of Eqs. (19)-(22). Further support to this interpretation could be given recalling the considerations on relativistic equations of motion for spin, see e. g. Section 11.11 in \cite{9}: Thomas precession of spin can be understood using P and T odd 4-vector $S^\mu$ that couples to the EM field tensor $F^{\mu\nu}$ in the form $S_\nu F^{\mu\nu}$ to give the equation of motion. Moreover note that discussing the physics of TMI \cite{5} authors argue that static axion field could become dynamical in the antiferromagnetic phase due to spin wave and amplitude excitations underlining the significance of spin interaction in this phenomenon.

Remarkably $W_\mu$ becomes axion-like field assuming ${\bf W} =-{\bf \nabla}a$ and $W_0 = -\frac{\partial a}{\partial t}$, and setting magnetic charge current zero when (19) to (22) reduce to
\begin{equation}
{\bf \nabla.D} = -g {\bf \nabla}a.{\bf B} +4\pi \rho_e 
\end{equation}
\begin{equation}
{\bf \nabla.B} = 0 
\end{equation}
\begin{equation}
{\bf \nabla} \times {\bf E} + \frac {\partial {\bf B}}{\partial t} - g \frac{\partial a}{\partial t} {\bf E} =0
\end{equation}
\begin{equation}
{\bf \nabla} \times {\bf H} - \frac {\partial {\bf D}}{\partial t} - g( {\bf \nabla}a \times {\bf E} +\frac{\partial a}{\partial t} {\bf B}) =4 \pi {\bf J}_e 
\end{equation}
To obtain above equations it is necessary that ${\bf \nabla}a$ is normal to ${\bf E}$ and parallel to ${\bf B}$: this is an important restriction that does not follow from the derivation of axion electrodynamics using the topological action (3) in \cite{2}. Further Faraday's law (25) also gets modified when $a$ is time-dependent as compared to axion electrodynamics in which Faraday's law is unaltered. What about the key ingredient: topological quantization? Though there is no topological action in the present formulation topological quantization arises if duality gauge field is not single-valued. It is similar to the topology of Aharonov-Bohm effect \cite{11}: in the standard illustration a multiply connected space follows for $a=\phi$ in cylindrical coordinates $(r, \phi, z)$ in which periodic boundary $\phi \rightarrow \phi +2\pi$ changes the field $a$. Once this topological quantization is used the arguments given by Qi, Hughes and Zhang in \cite{2} could be immediately adopted to arrive at the new properties of TI i. e. TME and image monopole, since the field equations (23) to (26) are identical to their equation (104) for static $a$. Interestingly the gauge coupling constant may also be obtained using multi-valuedness of $a$ and transforming $g {\bf \nabla}a.{\bf B}$ in (23) to $g\phi \Phi_0$ where $\Phi_0 =\frac{hc}{2e}$ is magnetic flux quantum: changing $a$ from 0 to $2\pi$ induces an electric charge, if it is assumed to be $4\pi e$ then we get $g=2\alpha /\pi$.

Preceding discussion shows that duality invariant theory explains all the novel features of topological insulators for static $a$, and predicts a modification in Eq. (25) for dynamical $a$. Expected consequence of modification in Eq. (25) would be that the dispersion relation of so called axionic polariton proposed in \cite{5} will change. Next we discuss an entirely new effect that emerges in the duality invariant theory.

Let us consider dual of axion-like electrodynamics: assume ${\bf \nabla}a$ normal to ${\bf B}$ and parallel to ${\bf E}$ then Eqs. (19) to (22) reduce to the following equations
\begin{equation}
 {\bf \nabla.D} = 4\pi \rho_e 
\end{equation}
\begin{equation}
 {\bf \nabla.B} =  g {\bf \nabla}a.{\bf E}
\end{equation}
\begin{equation}
 {\bf \nabla} \times {\bf E} + \frac {\partial {\bf B}}{\partial t} + g( {\bf \nabla}a \times {\bf B} +\frac{\partial a}{\partial t} {\bf E})=0
\end{equation}
\begin{equation}
 {\bf \nabla} \times {\bf H} - \frac {\partial {\bf D}}{\partial t} - g \frac{\partial a}{\partial t} {\bf B} =4 \pi {\bf J}_e 
\end{equation}
Eq. (28) shows that electric field induces a magnetic charge, and corresponding magnetic current in Eq. (29). In contrast to axion electrodynamics its dual is thus seen to be characterized by a purely duality (topological) electrodynamics as free magnetic source current and material parameters
$\epsilon$ and $\mu$ are not present in Eqs. (28) and (29). Note that axion field effect in TI is relatively very small as compared to free current $J^\mu_e$ and the parameters $\epsilon$ and $\mu$ and though one is looking for surface effects their observation becomes very difficult. Therefore search for possible materials and viable experiments to detect the dual effect embodied in Eqs. (27) to (30) would be of great value.

An intriguing aspect regarding induced magnetic charge is worth mentioning. To calculate it in Eq. (28) we perform the similar steps as we did for calculating induced electric charge in Eq. (23). Assuming electric flux equal to $4\pi e$ induced magnetic charge is obtained to be $16 \pi e \alpha$. Recalling that monopole charge according to Dirac quantization condition is $e/2\alpha$, the ratio of induced magnetic charge to monopole charge turns out to be extremely small: $32\pi \alpha ^2$.

In the light of symmetry and universality for TI electrodynamics emphasized in \cite{2} it becomes natural and attractive to explore generalized duality invariance. However it is imperative to seek its foundation or link with some variant of band theory that takes into consideration explicit calculation of various interactions in the material. Dynamical axion in MTI may have origin in the coupling of electrons with the antiferromagnetic order parameter \cite{6}, therefore it would be interesting to examine the role of charge density wave order parameter afresh in conjunction with ferromagnetic and antiferromagnetic order parameters in new materials if the dual to axion field predicted in the present work could be realized. In fact, discussing SL(2, Z) symmetry in the case of dynamical axion field Karch \cite{8} has rightly noted that duality demands $\epsilon$ and $\mu$ also to become dynamical fields - the dilaton. In our duality invariant theory, the complete set of Eqs. (15) and (16) shows that the material parameters $\epsilon$ and $\mu$ need not be dynamical variables rather the dynamical axion field is a manifestation of the duality gauge field.

Before we conclude, the significance of topological phases in optics, in particular, Pancharatnam phase for unraveling the unique properties of TI, for example, TME is pointed out. Though optical techniques have been extensively used to study interfaces and surface properties of insulators a recent experiment \cite{12} seems important in which TI is probed using ultrafast laser pulses in the nonlinear regime. Duality invariant rendition of TI electrodynamics presented here indicates the role of spin angular momentum and topological phase of polarized light besides the topological property of the insulating material. We suggest that the kind of experiment performed by Hsieh et al \cite{12} could be enlarged in scope to search for any such topological phase effect in TI. One possibility would be to study optical rotation of polarized light incident on vacuum - TI - ordinary insulator and observe reflected and transmitted light for optical rotation \cite{2, 8}, if any, and analyze the observations for Pancharatnman phase.

In conclusion a generalized duality invariant electrodynamics is developed applicable to topological insulators and it is shown that under appropriate conditions this theory reduces to the axion field-like theory widely discussed in the literature; a new effect dual to axion electrodynamics is also predicted.

Library facility at Banaras Hindu University, Varanasi is acknowledged.


\begin{thebibliography}{99}
\bibitem{1} L. I. Schiff, Quantum Mechanics (McGraw-Hill, third edition,1968)
\bibitem{2} X-L Qi, T. L. Hughes and S-C Zhang, Phys. Rev. B 78, 195424 (2008); X-L Qi and S-C Zhang, Phys. Today 63(1), 33 (2010)
\bibitem{3} A. M. Essin, J. E. Moore and D. Vanderbilt, Phys. Rev. Lett. 102, 146805 (2009); M. Z. Hasan and C. L. Kane, Rev. Mod. Phys. 82, 3045 (2010); M. Z. Hasan and J. E. Moore, arXiv: 1011.5462
\bibitem{4} X-L Qi, Q Li, J. Zang and S-C Zhang, Science, 323, 1184 (2009)
\bibitem{5} R. Li, J. Wang, X Qi and S-C Zhang, arXiv: 0908.1537 v1
\bibitem{6} F. Wilczek, Phys. Rev. Lett. 58, 1799 (1987)
\bibitem{7} E. Witten, Phys. Today 50(5), 28 (1997)
\bibitem{8} A. Karch, Phys. Rev. Lett. 103, 171601 (2009)
\bibitem{9} J. D. Jackson, Classical Electrodynamics (Second Edition, John Wiley, 1975)
\bibitem{10} A. Sudbery, J. Phys. A Math. Gen. 19, L33 (1986); S. C. Tiwari, Physics Essays 4, 212 (1991)
\bibitem{11} L. H. Ryder, Quantum Field Theory (C. U. P. 1985)
\bibitem{12} D. Hsieh, J. W. McIver, D. H. Torchinsky, D. R. Gardner, Y. S. Lee and N. Gedik, Phys. Rev. Lett 106, 057401 (2011)
 
\end{thebibliography}
\end{document}